\documentclass{amsart}
\usepackage[utf8]{inputenc}
\usepackage[foot]{amsaddr}
\usepackage{fullpage}

\usepackage{amsmath,xcolor,diagbox,tikz,float}
\usepackage{amssymb}
\usepackage[normalem]{ulem}
\usepackage{epstopdf}
\usepackage{mathrsfs}
\usepackage{soul}
\usepackage{bm}
\usepackage{comment}
\usepackage{mathtools}
\usepackage{optidef}
\usepackage{booktabs}
\usepackage{url}
\usepackage{xcolor}
\usepackage[caption=false, font=footnotesize]{subfig}
\def\BibTeX{{\rm B\kern-.05em{\sc i\kern-.025em b}\kern-.08em
    T\kern-.1667em\lower.7ex\hbox{E}\kern-.125emX}}
    
\ifpdf
  \DeclareGraphicsExtensions{.eps,.pdf,.png,.jpg}
\else
  \DeclareGraphicsExtensions{.eps}
\fi
\newcommand{\CE}{\mathcal{E}}

\newcommand{\CV}{\mathcal{V}}

\DeclareMathOperator*{\tr}{tr}

\newcommand{\Norm}[2][]{\lVert#2\rVert\ifthenelse{\isempty{#1}}{}{_{#1}}}
\newcommand{\BigNorm}[2][]{\left\lVert#2\right\rVert\ifthenelse{\isempty{#1}}{}{_{#1}}}
\newcommand{\InProd}[3][]{\langle#2,#3\rangle\ifthenelse{\isempty{#1}}{}{_{#1}}}
\newcommand{\BigInProd}[3][]{\left\langle#2,#3\right\rangle\ifthenelse{\isempty{#1}}{}{_{#1}}}

\newcommand{\yzcmt}[1]{\textcolor{black}{#1}}
\begin{document}

\author{Yabin Zhang$^{1}$}
\author{David Gorsich$^{2}$}
\author{Paramsothy Jayakumar$^{2}$}
\author{Shravan Veerapaneni$^{1*}$}
\address{$^1$Department of Mathematics, University of Michigan, Ann Arbor, MI 48109 USA}
\address{$^2$Ground Vehicle Systems Center, U.S. Army DEVCOM, Warren, MI}
\thanks{$^*$Corresponding author. \emph{Email address:} shravan@umich.edu \,}

 \title{Continuous-variable optimization with neural network quantum states}

\maketitle

\begin{abstract}
Inspired by proposals for continuous-variable quantum approximate optimization (CV-QAOA), we investigate the utility of continuous-variable neural network quantum states (CV-NQS) for performing continuous optimization, focusing on the ground state optimization of the classical antiferromagnetic rotor model.
Numerical experiments conducted using variational Monte Carlo with CV-NQS indicate that although the non-local algorithm succeeds in finding ground states competitive with the local gradient search methods, the proposal suffers from unfavorable scaling. A number of proposed extensions are put forward which may help alleviate the scaling difficulty.

\end{abstract}

\section{Introduction} %
It is an intriguing fact that many computationally difficult combinatorial optimization problems can be encoded in the ground states of associated physical systems. Classic examples include the {\em Max-Cut} and {\em Maximal Independent Set} problems defined on $n$ bits, whose solution sets correspond to the minimal energy ground state configurations for the classical anti-ferromagnetic Ising model and hard-core gas model, respectively. The existence of identifications between combinatorial solution sets and ground states has motivated the pursuit of physically-inspired heuristic approximation algorithms. Perhaps the  most famous example of this kind is simulated annealing (SA) \cite{kirkpatrick1983optimization}, which attempts to interpolate, via a sequence of approximate Gibbs distributions, between the uniform distribution on $n$ bits and the uniform distribution on the solution set. Likewise, by  encoding the cost function on $n$ bits as the endpoint of a one-parameter family of quantum Hamiltonians, quantum annealing (QA) \cite{farhi2000quantum} attempts to follow a trajectory in Hilbert space, which interpolates, via a sequence of approximate ground states, between the uniform superposition state of $n$ qubits and a state in the linear span of orthonormal basis elements corresponding to valid solutions. More recently, partially inspired by the search for practical utility of noisy intermediate-scale quantum computers, variational implementations  of both QA and SA have been advocated, which have been termed quantum approximate optimization algorithm (QAOA) \cite{farhi2014quantum} and variational neural annealing (VNA) \cite{hibat2021variational}. These variational algorithms achieve the desired interpolation by optimizing over a space of trial wavefunctions (respectively, probability distributions), which are selected from a variational class by following the gradient of a stochastic objective function,  estimated via Monte Carlo sampling. Both SA and VNA can thus be viewed as instances of a generalized variational Monte Carlo (VMC) with discrete classical configuration space. 

Simulated annealing for continuous optimization has been investigated \cite{nikolaev2010simulated} and continuous variants of QAOA (so called CV-QAOA) have also been recently introduced \cite{verdon2019quantum} in the context of continuous-variable quantum computing. This paper arose from a desire to understand the contexts in which CV-QAOA might offer an advantage over classical local update algorithms such as gradient descent. Motivated by this goal, we consider a class of NP-hard continuous optimization problems from two perspectives. The first is a gradient based Riemannian trust-region method due to Burer-Monteiro-Zhang (BMZ) \cite{BMZ2001}. The second is a non-local probabilistic algorithm inspired by CV-QAOA. Due to the inability to efficiently simulate CV-QAOA at scale, we replace it by a simulation strategy based on neural network quantum states \cite{stokes2021continuous}. In particular, the variational wavefunction is chosen to be a rotor variant of the restricted Boltzmann machine, optimized via stochastic natural gradient descent \yzcmt{\cite{carleo2017solving,Amari1998,zhao2020natural}} using a random-walk Metropolis-Hastings Markov-Chain Monte Carlo strategy. The use of restricted Boltzmann machines (RBMs) enables for an efficient initialization strategy using the output of the BMZ algorithm. The RBM suffers the disadvantage of depending upon an unknown normalizing constant, which precludes the possibility of utilizing an annealing schedule as discussed in \cite{hibat2021variational}. In this preliminary work, we therefore focus on the limit of zero entropy regularization in which the variational Monte Carlo reduces to natural evolution strategies as discussed in \yzcmt{\cite{gomes2019classical, zhao2020natural}}.
\yzcmt{The second approach with neural quantum states (NQS) is quantum-inspired since the considered Hamiltonian for the optimization problem can be regarded as that for a system of quantum oscillators with zero kinetic term. This approach is based upon previous work \cite{carleo2017solving} and more recent work \cite{stokes2021continuous} which use neural-network quantum states to simulate quantum systems. }

The paper is structured as follows. In section \ref{sec:background} we introduce the continuous optimization problem, explaining its relationship with semi-definite program relaxations for Max-Cut and the mean-field description of quantum anti-ferromagnets. Next we outline the proposed non-local optimization algorithm based on neural network trial densities. The numerical results presented in section \ref{s:results} demonstrate the ability of the algorithm to accelerate using BMZ initialization. The exploration of different regularization strategies and quantum extensions is left to future work.

\section{Background}\label{sec:background}
In the computer science literature, a successful strategy for developing approximation heuristics, as well as approximation algorithms with provable approximation guarantees, involves the concept of relaxation of the search space. In the case of Max-Cut, a celebrated approximation algorithm due to Goemans and Williamson \cite{goemans1995improved}, proceeds by enlarging the binary search space over $n$ bits, to the product manifold $\prod_{i=1}^n \mathbb{S}^{n-1}$, where $\mathbb{S}^{n-1}$ denotes the unit sphere of dimension $n-1$, isometrically embedded in $n$-dimensional Euclidean space $\mathbb{R}^n$. The cost function defined on the enlarged search space is equivalent to the energy of the classical antiferromagnetic $O(n)$-invariant rotor model, in which the spins are $n$-dimensional unit vectors. A configuration of rotors achieving the optimal ground state energy can be efficiently found using semidefinite programming and can be mapped by a randomized procedure to a classical bit string satisfying an approximation guarantee. In practice, however, heuristic approximation algorithms are found to significantly outperform the Goemans-Williamson approximation guarantee. In particular, the BMZ algorithm \cite{BMZ2001}, replaces the rank-$n$ relaxation step of Goemans and Williamson with rank-2 relaxation, resulting in the configuration space of $n$ planar rotors $\prod_{i=1}^n \mathbb{S}^2$ with energy given by the antiferromagnetic $O(2)$-invariant rotor model. Although global energy optimization of the planar rotor model is not amenable to semidefinite programming, local energy optimization can be efficiently performed and BMZ produces superior cut values to GW in practice.

In order to fix notation let $G=(\CV, \CE)$ be a simple undirected graph with vertex set $\CV = \{1, \ldots, n\}$ for some $n \geq 2$ \yzcmt{and edge set $\CE$}. In addition, let the weight for the edge between vertex $i$ and $j$ be $w_{ij}$.
\yzcmt{Note that $w_{ij}=0$ for any pair of $(i,j)\notin \CE$  and $w_{ij}=w_{ji}$ for any $i,j\in\CV$.}

\subsection{Relaxation of Max-Cut to rotor model}
\label{s:relax}
The Max-Cut problem defined on $G$ finds a bipartition of the graph such that the total weights on the edges  between the two bipartite sets is maximized.
The Max-Cut problem can be formulated as a binary optimization problem
\begin{mini}|l|
  {}{\frac{1}{2}\sum_{\{i,j\} \in \mathcal{E}} w_{ij}(1-x_i x_j)}{}{}
  \addConstraint{ \lvert x_i \lvert }{=1,}{\quad i=1,\ldots,n}
  \label{eq:binary_maxcut1}
\end{mini}
Since the Max-Cut problem is NP-hard, we follow \cite{BMZ2001} by considering
a \yzcmt{rank-2} relaxation together with a heuristic algorithm.
This relaxation replaces $x_i\in \mathbb{R}$ with $\vec{v}_i\in\mathbb{R}^2$, yielding
\begin{mini}|l|
  {}{\sum_{\{i,j\} \in \mathcal{E}} w_{ij} \langle \vec{v}_i, \vec{v}_j \rangle}{}{}
  \addConstraint{ \Vert \vec{v}_i \Vert }{=1,}{\quad i=1,\ldots,n}
\end{mini}
Since $v_i$ lives on the unit circle in $\mathbb{R}^2$, the above minimization problem is equivalent to the unconstrained minimization problem by switching to polar coordinates, where $\theta_i\in\mathbb{R}$,
\begin{mini}|l|
  {}{\sum_{\{i,j\} \in \mathcal{E}} w_{ij} \cos(\theta_i - \theta_j)}{}{}
  \label{eq:relaxation_angle_maxcut}
\end{mini}

A solution to the original Max-Cut problem can be obtained by
first solving the unconstrained optimization problem for $\theta=(\theta_1,\dots,\theta_n)$ and then applying the heuristic algorithm ``Procedure-Cut'' in \cite{BMZ2001} which maps the angle $\theta$ to a valid cut $x=(x_1,\dots,x_n)$ for the graph. 

\subsection{Heisenberg model}
As additional motivation for the continuous optimization problem \eqref{eq:relaxation_angle_maxcut}, consider the following Heisenberg quantum Hamiltonian formulated on the graph $G$,
\begin{equation}
    H = \sum_{\{i,j\} \in \mathcal{E}} w_{ij} (X_i X_j + Z_i Z_j) \enspace .
\end{equation}
and define the following variational wavefunction parametrized by $\theta = (\theta_1,\ldots,\theta_n)$,
\begin{equation}
    \rho = \bigotimes_{i=1}^n \frac{1}{2}(I + \sin\theta_i X_i + \cos\theta_i Z_i) \enspace .
\end{equation}
Evaluating the quantum expectation value of $H$ in the state $\rho$ gives the variational upper bound $\tr(H \rho) \geq \lambda_{\min}(H)$ for all $\theta$. Optimizing the variational bound over $\theta$ is equivalent to problem \eqref{eq:relaxation_angle_maxcut}.
\section{Neural quantum state relaxation}
In this section we pass from deterministic rotor configurations $\theta \in [0,2\pi)^n$ to parametrized probability densities over $[0,2\pi)^n$ satisfying periodic boundary conditions. In particular, we introduce a set of $m$ hidden rotor variables $\varphi = (\varphi_1,\ldots,\varphi_m) \in [0,2\pi)^m$ and define a wavefunction by integrating a Boltzmann factor with respect to the flat measure,
\begin{equation}
    \psi(\theta) =
    \int_{[0,2\pi]^m} {\rm d}^m\varphi \,
    \exp\big({-}E(\theta,\varphi)\big)
\end{equation}
where the energy is of the following restricted Boltzmann form
\begin{equation}
    E(\theta,\varphi) = -\sum_{i=1}^m\sum_{j=1}^n a_{ij} \langle \vec{z}_i, \vec{v}_j \rangle  - \sum_{i=1}^m \langle \vec{b}_i, \vec{z}_i \rangle - \sum_{j=1}^n \langle \vec{c}_j, \vec{v}_j \rangle \enspace.
\end{equation}
Here  we have passed back to Cartesian coordinates $\vec{v}_i = (\cos\theta_i, \sin\theta_i)$ and $\vec{z}_i = (\cos\varphi_i,\sin\varphi_i)$. The wavefunction is parametrized by weights $a_{ij} \in \mathbb{R}$ and biases $\vec{b}_i \in \mathbb{R}^2$ and $\vec{c}_j \in \mathbb{R}^2$ for all $(i,j) \in [m] \times [n]$. It is equally possible to consider holomorphic parametrization in the spirit of \cite{carleo2017solving}, although we only consider real parametrization in this work. The wavefunction defines a probability density given by the Born rule,
\begin{equation}
    \pi(\theta) := \frac{\lvert \psi(\theta) \lvert^2}{Z}
\end{equation}
which is normalized by the partition function
\begin{equation}
    Z := \int_{[0,2\pi]^n} {\rm d}^n\theta \, \lvert\psi(\theta) \lvert^2 \enspace .
\end{equation} Although the integral defining the partition function is intractable, the integrals over $\varphi$ can be computed in closed form owing to the bipartite structure of the energy function. It follows that the probability density $\pi(\theta)$ over visible rotor configurations is known up to an overall normalizing constant. The stochastic optimization problem is thus defined as
\begin{mini}|l|
  {}{\underset{\theta \sim \pi}{\mathbb{E}} \left[\sum_{\{i,j\} \in \mathcal{E}} w_{ij} \cos(\theta_i - \theta_j)\right]}{}{}
  \label{eq:relaxation_density}
\end{mini}
where the unconstrained minimization is over the density class parameterized by $\mathbb{R}^{nm + 2(n+m)}$. In practice, this minimization is performed using stochastic reconfiguration \yzcmt{\cite{mcmillan1965ground, carleo2017solving,Amari1998}}. \yzcmt{For details on the stochastic configuration, please refer to the supplemental materials of \cite{carleo2017solving}.}


\begin{figure}
\begin{center}
\scalebox{0.75}{\tikzset{every picture/.style={line width=0.75pt}} 

\begin{tikzpicture}[x=0.75pt,y=0.75pt,yscale=-1,xscale=1]

\draw  [color={rgb, 255:red, 128; green, 128; blue, 128 }  ,draw opacity=1 ][line width=1.5]  (152,37) .. controls (152,30.92) and (156.92,26) .. (163,26) -- (343,26) .. controls (349.08,26) and (354,30.92) .. (354,37) -- (354,70) .. controls (354,76.08) and (349.08,81) .. (343,81) -- (163,81) .. controls (156.92,81) and (152,76.08) .. (152,70) -- cycle ;
\draw  [color={rgb, 255:red, 128; green, 128; blue, 128 }  ,draw opacity=1 ][line width=1.5]  (152,162.8) .. controls (152,156.84) and (156.84,152) .. (162.8,152) -- (334.2,152) .. controls (340.16,152) and (345,156.84) .. (345,162.8) -- (345,195.2) .. controls (345,201.16) and (340.16,206) .. (334.2,206) -- (162.8,206) .. controls (156.84,206) and (152,201.16) .. (152,195.2) -- cycle ;
\draw  [color={rgb, 255:red, 128; green, 128; blue, 128 }  ,draw opacity=1 ][line width=1.5]  (230,429) .. controls (230,409.12) and (281.49,393) .. (345,393) .. controls (408.51,393) and (460,409.12) .. (460,429) .. controls (460,448.88) and (408.51,465) .. (345,465) .. controls (281.49,465) and (230,448.88) .. (230,429) -- cycle ;
\draw  [color={rgb, 255:red, 128; green, 128; blue, 128 }  ,draw opacity=1 ][line width=1.5]  (284,285.8) .. controls (284,279.84) and (288.84,275) .. (294.8,275) -- (454.2,275) .. controls (460.16,275) and (465,279.84) .. (465,285.8) -- (465,318.2) .. controls (465,324.16) and (460.16,329) .. (454.2,329) -- (294.8,329) .. controls (288.84,329) and (284,324.16) .. (284,318.2) -- cycle ;
\draw  [color={rgb, 255:red, 128; green, 128; blue, 128 }  ,draw opacity=1 ][line width=1.5]  (164,528.5) .. controls (164,514.42) and (198.92,503) .. (242,503) .. controls (285.08,503) and (320,514.42) .. (320,528.5) .. controls (320,542.58) and (285.08,554) .. (242,554) .. controls (198.92,554) and (164,542.58) .. (164,528.5) -- cycle ;
\draw [color={rgb, 255:red, 28; green, 10; blue, 238 }  ,draw opacity=1 ] [dash pattern={on 4.5pt off 4.5pt}]  (221.19,214.92) .. controls (200.69,241.41) and (189.01,270.11) .. (183.25,298.81) .. controls (180.01,314.98) and (178.64,331.15) .. (178.64,346.93) .. controls (178.64,388.84) and (188.26,427.98) .. (197.87,457.01) .. controls (203.27,473.31) and (208.67,486.43) .. (210.09,489.67)(218.81,213.08) .. controls (198.01,239.97) and (186.15,269.09) .. (180.31,298.22) .. controls (177.03,314.58) and (175.64,330.95) .. (175.64,346.93) .. controls (175.64,389.19) and (185.33,428.67) .. (195.03,457.95) .. controls (200.46,474.36) and (205.89,487.56) .. (207.32,490.81) ;
\draw [shift={(212,498)}, rotate = 246.61] [fill={rgb, 255:red, 28; green, 10; blue, 238 }  ,fill opacity=1 ][line width=0.08]  [draw opacity=0] (14.29,-6.86) -- (0,0) -- (14.29,6.86) -- cycle    ;
\draw    (263.49,211.58) .. controls (291.98,221.41) and (317.6,241.73) .. (328.34,262.87)(262.51,214.42) .. controls (290.28,223.99) and (315.27,243.77) .. (325.74,264.38) ;
\draw [shift={(331,271)}, rotate = 244.29000000000002] [fill={rgb, 255:red, 0; green, 0; blue, 0 }  ][line width=0.08]  [draw opacity=0] (14.29,-6.86) -- (0,0) -- (14.29,6.86) -- cycle    ;
\draw    (356.3,338.25) .. controls (364.28,352.04) and (366.08,363.77) .. (366.08,375.27) .. controls (366.08,376.69) and (366.05,378.11) .. (366,379.53) .. controls (365.93,381.75) and (365.81,383.98) .. (365.98,380.23)(353.7,339.75) .. controls (361.36,352.99) and (363.08,364.24) .. (363.08,375.27) .. controls (363.08,376.66) and (363.05,378.05) .. (363,379.43) .. controls (362.93,381.63) and (362.81,383.83) .. (362.98,380.06) ;
\draw [shift={(364,389)}, rotate = 273.58] [fill={rgb, 255:red, 0; green, 0; blue, 0 }  ][line width=0.08]  [draw opacity=0] (14.29,-6.86) -- (0,0) -- (14.29,6.86) -- cycle    ;
\draw    (353.14,469.98) .. controls (342.82,482.01) and (339.53,486.46) .. (336.67,489.94) .. controls (336.48,490.17) and (336.3,490.39) .. (336.11,490.61) .. controls (335.87,490.9) and (335.63,491.18) .. (335.38,491.47) .. controls (332.1,495.25) and (328.11,499.04) .. (318.18,508.33)(350.86,468.02) .. controls (340.52,480.09) and (337.23,484.55) .. (334.35,488.03) .. controls (334.17,488.25) and (334,488.46) .. (333.82,488.68) .. controls (333.58,488.95) and (333.35,489.23) .. (333.11,489.5) .. controls (329.87,493.25) and (325.92,496.99) .. (316.09,506.18) ;
\draw [shift={(311,513)}, rotate = 316.90999999999997] [fill={rgb, 255:red, 0; green, 0; blue, 0 }  ][line width=0.08]  [draw opacity=0] (14.29,-6.86) -- (0,0) -- (14.29,6.86) -- cycle    ;
\draw  [color={rgb, 255:red, 128; green, 128; blue, 128 }  ,draw opacity=1 ][line width=1.5]  (170,633) .. controls (170,621.4) and (201.56,612) .. (240.5,612) .. controls (279.44,612) and (311,621.4) .. (311,633) .. controls (311,644.6) and (279.44,654) .. (240.5,654) .. controls (201.56,654) and (170,644.6) .. (170,633) -- cycle ;
\draw    (250.5,91) -- (250.5,138)(247.5,91) -- (247.5,138) ;
\draw [shift={(249,147)}, rotate = 270] [fill={rgb, 255:red, 0; green, 0; blue, 0 }  ][line width=0.08]  [draw opacity=0] (14.29,-6.86) -- (0,0) -- (14.29,6.86) -- cycle    ;
\draw    (241.5,556) -- (241.5,585.96) -- (242.07,597.94)(238.5,556) -- (238.5,586.04) -- (239.07,598.08) ;
\draw [shift={(241,607)}, rotate = 267.27] [fill={rgb, 255:red, 0; green, 0; blue, 0 }  ][line width=0.08]  [draw opacity=0] (14.29,-6.86) -- (0,0) -- (14.29,6.86) -- cycle    ;

\draw (166.51,37.9) node [anchor=north west][inner sep=0.75pt]  [rotate=-359.77] [align=center] {Max-Cut as constrained \\binary minimization problem};
\draw (132,99) node [anchor=north west][inner sep=0.75pt]   [align=center] {Rank-2 \ \\relaxation in \cite{BMZ2001}\\};
\draw (259.13,99.73) node [anchor=north west][inner sep=0.75pt]  [rotate=-0.34] [align=center] {Switch to polar\\coordinates};
\draw (170,164) node [anchor=north west][inner sep=0.75pt]   [align=center] {Unconstrained continuous \\minimization problem};
\draw (333.05,220.8) node [anchor=north west][inner sep=0.75pt]  [rotate=-359.93] [align=center] {Neural quantum \\state relaxation};
\draw (305,287) node [anchor=north west][inner sep=0.75pt]   [align=center] {Stochastic minimization \\problem};
\draw (266,415) node [anchor=north west][inner sep=0.75pt]   [align=center] {Approximate wavefunction\\probability density };
\draw (379.92,347.19) node [anchor=north west][inner sep=0.75pt]  [rotate=-359.64] [align=center] {Stochastic \\reconfiguration};
\draw (182,522) node [anchor=north west][inner sep=0.75pt]   [align=center] {Rotor configuration};
\draw (346.04,482.89) node [anchor=north west][inner sep=0.75pt]  [rotate=-0.19] [align=center] {Sampling};
\draw (198,627) node [anchor=north west][inner sep=0.75pt]   [align=center] {Cut on graph};
\draw (255.71,561.7) node [anchor=north west][inner sep=0.75pt]  [rotate=-359.23] [align=center] {Procedure-cut \\algorithm in \cite{BMZ2001}};
\draw (75,329) node [anchor=north west][inner sep=0.75pt]   [align=center] {\textcolor[rgb]{0.04,0.02,0.96}{Deterministic }\\\textcolor[rgb]{0.04,0.02,0.96}{minimization }\\\textcolor[rgb]{0.04,0.02,0.96}{algorithm, }\\\textcolor[rgb]{0.04,0.02,0.96}{e.g. trust }\\\textcolor[rgb]{0.04,0.02,0.96}{region methods}};

\end{tikzpicture}
}

\caption{Flow chart for solving the Max-Cut problem: the (black) solid arrows represent the NQS algorithm proposed in this manuscript; the (blue) dashed line arrow represent an deterministic approach given in \cite{BMZ2001}.   }
\label{fig:flow_chart_relaxation}
\end{center}
\end{figure}
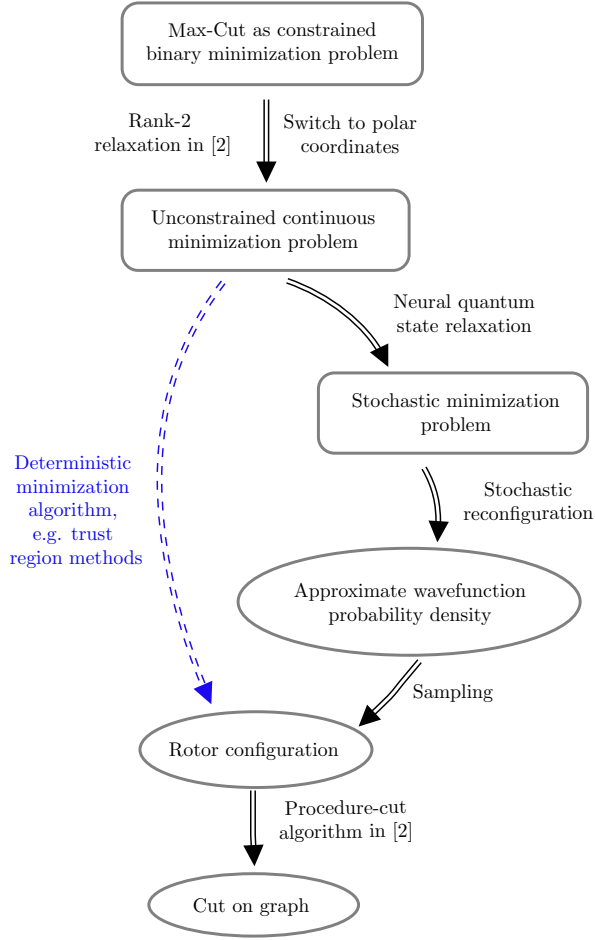


\section{Numerical results}
\label{s:results}

We investigated the performance of the NQS solver 
on two types of graphs, characterized by having either randomized or uniform edge weights, respectively. 
Previous studies on Kuramoto oscillators show that the continuous optimization problem (\ref{eq:relaxation_angle_maxcut}) 
formulated with uniform weights and high edge density can be solved via local algorithms since all local extremas are also global extremas \cite{Steinerberger2021MaxCutVK}.  
Since the BMZ algorithm uses only local information whereas the NQS algorithm updates a global density,
the latter potentially holds promise in producing
improved solutions in the regime where local extremas are not always global extremas.
Thus, we included some low edge density uniform weight graphs in our tests in addition to general random weight graphs.

We will consider the following two graphs in particular:
\begin{itemize}
\item \texttt{RWN50E619}: a graph with 50 nodes and 619 edges with random weight for each edge in the range (0,15). The edge density is approximately 50\%. 
\item \texttt{UWN100E519}: a graph with 100 nodes and 519 edges with  uniform weight for each edge. The edge density is about 10\%.
\end{itemize}

For comparison with NQS we use an implementation of the BMZ algorithm in  \cite{BMZ2001}.
\yzcmt{The main idea of the BMZ approach is to relax the original binary minimization problem of Max-cut to a unconstrained continuous variable minimization problem defined on $\mathbb{R}^n$, where $n$ is the number of nodes on the graph. The Hessian matrix of the relaxed problem is easy to compute and thus the relaxed problem can be solved via standard classical minimization algorithms such as the trust-region method used in our implementation.}
The solution results for the two tested graphs are given in Table \ref{tab:bmz_twographs}.
\begin{table}
\begin{center}
\begin{tabular}{@{}c|cccc@{}}
\toprule
 & 
\multicolumn{3}{c}{Energy value}\\
Graph& Mean & Standard deviation & Minimum\\
 \midrule
\texttt{RWN50E619} &  -1449.35&	10.5&   -1459.35 \\
\texttt{UWN100E519} & -168.45&	1.52&  -170.65\\
 \bottomrule
\end{tabular}
\end{center}
\caption{Observed solution energy for applying the BMZ algorithm to the two tested graphs.}
\label{tab:bmz_twographs}
\end{table}

The proposed NQS algorithm has several parameters whose values were found to control the solution quality. 
Some of these parameters are held fixed throughout all testing. In particular,
\yzcmt{the hidden unit density, i.e., the ratio between the number of hidden nodes and the visible nodes $\alpha=m/n$,} was set to $1$, the descent direction was solved via MINRES with TOL=$10^{-10}$, and the learning rate was set to 0.01. 
Parameters that were varied amongst the tests include 
\begin{itemize}
\item $N_{\rm samp}$ and $N_{\rm warm}$: the number of Metropolis samples and warm samples (early samples that are discarded when calculating the expectation) used in the Markov chain.
\item $N_{\rm iter}$: the number of stochastic reconfiguration steps
\item $\lambda_{\rm reg}$: the regularization parameter used in solving the linear system for each stochastic reconfiguration step
\end{itemize}
Section \ref{sec:RWgraph_parameter} describes how these parameters affect the solution for the \texttt{RWN50E619} graph. 
Both the BMZ algorithm and the proposed NQS algorithm
were executed 10 times with 10
random seeds, and the resulting mean energy, standard deviation, and minimum  energy are reported for all tests. 
\yzcmt{We focus on the solution quality rather than computational cost in  wall-clock time when comparing the results from the BMZ and the NQS approaches. As for running time, our implementation of BMZ solves the test problems on the order of 1 second while the NQS may require running time on the order of $10^{3}$ seconds. 
Such gap is expected as the sampling step of the NQS approach is costly and similar gap is also observed in previous work \cite{zhao2020natural}. }

Section \ref{sec:UWgraph_parameter} analyzes the NQS algorithm performance for uniform weight graphs with low edge density. 
While in theory we expect the proposed algorithm to outperform
the BMZ algorithm due to the latter's local nature, in practice sub-optimal solutions are found. 
The quality of the solutions is expected to improve with improved parameter tuning or simply by increasing $N_{\rm samp}$, $N_{\rm warm}$, and $N_{\rm iter}$. 

Finally, section \ref{sec:smart_initialization} discusses an initialization technique for the NQS algorithm, 
which exploits the output of the BMZ algorithm to initialize the weights and biases of the restricted Boltzmann machine. 

All tests are run on the Great Lakes cluster provided by Advanced Research Computing at the University of Michigan, with four 2x 3.0 GHz Intel Xeon Gold 6154 CPUs and 1GB memory per CPU requested per task. 
\yzcmt{Although this section is dedicated to test problems which require nontrivial optimization effort,
a collection of much smaller and/or simpler graphs are also included in the Appendix, which
verifies that the proposed approach produce the optimal solution for simple graphs.}

\subsection{Performance on random weight graph}\label{sec:RWgraph_parameter}
For the \texttt{RWN50E619} graph, 
the BMZ algorithm with 10 different random seeds produced an mean energy of -1449.35, standard deviation of 10.5, and minimum of -1459.35. 
In this section, we study how the VMC parameters in the NQS algorithm affect the solution quality when applied to this graph.

The impact of number of stochastic reconfiguration steps $N_{\rm iter}$ 
holding fixed $N_{\rm samp}=40, N_{\rm warm}=0, \lambda_{\rm reg}=10^{-9}$ is shown in Table \ref{tab:N50E619_Niter}. 
As $N_{\rm iter}$ increases, the energy approaches but is less than the value given by the BMZ algorithm.
\begin{table}
\begin{center}
\begin{tabular}{@{}c|cccc@{}}
\toprule
 & 
\multicolumn{3}{c}{Energy value}\\
$N_{\rm iter}$& Mean & Standard deviation & Minimum\\
 \midrule
8000  & -1359.29&	36.23&  -1405.88 \\
12000 & -1388.47&	 24.81&  -1416.89\\
16000 & -1404.48&	18.64&  -1431.56\\
20000 & -1407.50&	22.08&  -1435.02\\
 \bottomrule
\end{tabular}
\end{center}
\caption{Observed solution energy for the \texttt{RWN50E619} graph with respect to $N_{\rm iter}$, the number of stochastic reconfiguration steps. 
\yzcmt{$N_{\rm samp}=40$, $N_{\rm warm}=0$, and $\lambda_{\rm reg}=10^{-9}$ for all test cases.}
\yzcmt{See Table \ref{tab:bmz_twographs} for the corresponding results obtained by the BMZ algorithm.} 
}
\label{tab:N50E619_Niter}
\end{table}

\begin{figure}
\begin{center}
\scalebox{0.58}{\tikzset{every picture/.style={line width=0.75pt}} 


\begin{tikzpicture}[x=0.75pt,y=0.75pt,yscale=-1,xscale=1]
\Large 
\draw (388.5,273.75) node  {\includegraphics[width=401.25pt,height=278.63pt]{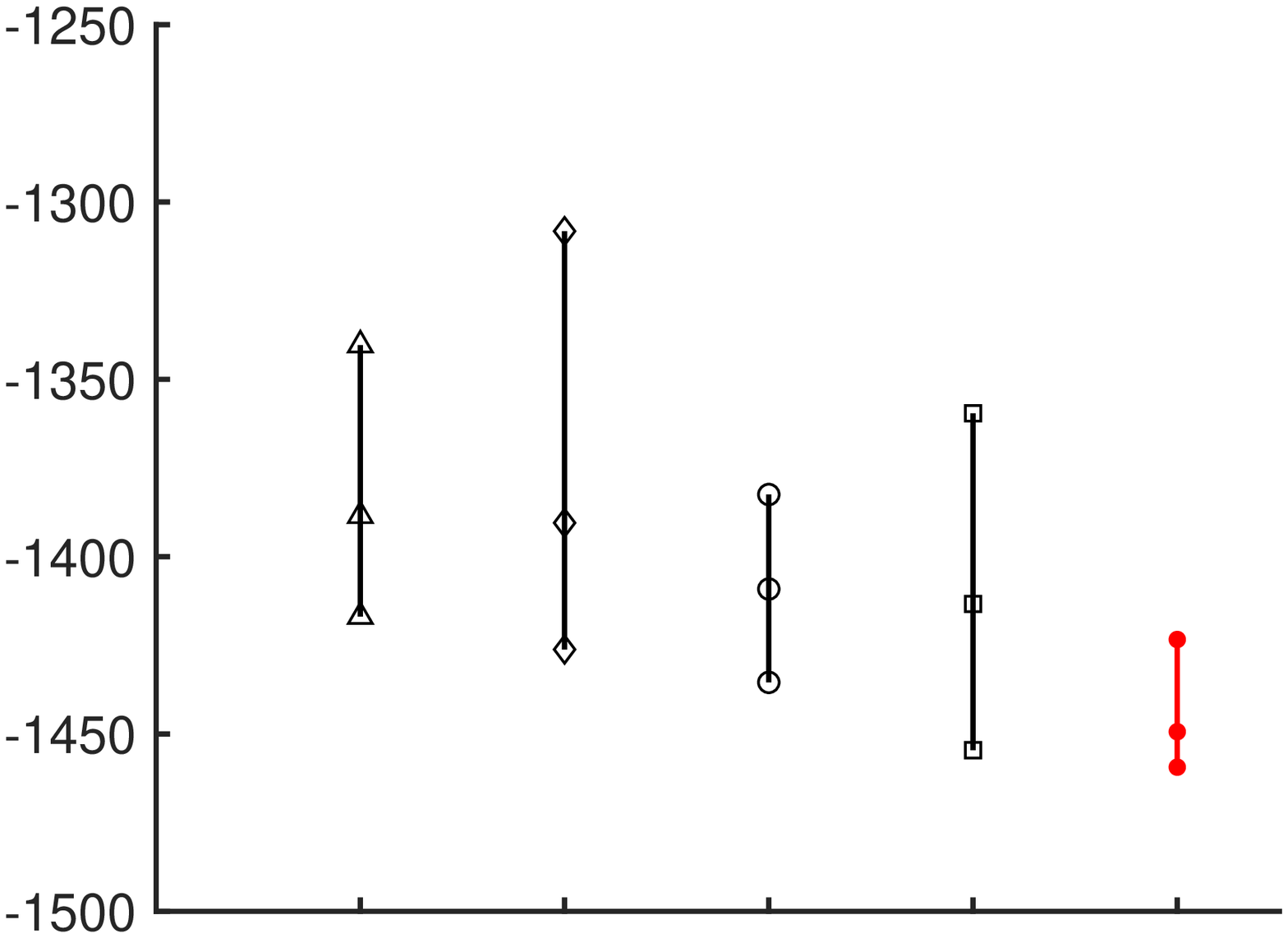}};

\draw (223.15,448.08) node [anchor=north west][inner sep=0.75pt]  [rotate=-333.59]  {$(40,0)$};
\draw (295.15,449.08) node [anchor=north west][inner sep=0.75pt]  [rotate=-333.59]  {$(100,10)$};
\draw (380.15,449.08) node [anchor=north west][inner sep=0.75pt]  [rotate=-333.59]  {$(200,20)$};
\draw (461.15,450.08) node [anchor=north west][inner sep=0.75pt]  [rotate=-333.59]  {$(400,40)$};
\draw (524,450.08) node [anchor=north west][inner sep=0.75pt]  [rotate=-333.59]  {\textcolor{red}{BMZ}};
\draw (337.88,474.16) node [anchor=north west][inner sep=0.75pt]  [rotate=-360]  {$(N_{\rm samp}, N_{\rm warm})$};
\draw (110.43,299.47) node [anchor=north west][inner sep=0.75pt]  [rotate=-270.14] [align=left] { Energy};

\end{tikzpicture}

}
\caption{Optimal energy found for the \texttt{RWN50E619} graph with respect to $N_{\rm samp}$ and $N_{\rm warm}$ the number of samples and warm samples in Markov chain. 
Each vertical line corresponds to one choice of the parameter pair $(N_{s\rm amp}, N_{\rm warm})$ or the BMZ algorithm,
and the three markers for each line give the observed minimum, average, and maximum of the energy approximations for 10 runs. \yzcmt{For the NQS approach,  $N_{\rm iter}=12000$ and $\lambda_{\rm reg}=10^{-9}$ for all test cases.} }
\label{fig:boxplotN50E619_Nsamp}
\end{center}
\end{figure}

We then fix $ \lambda_{\rm reg}=10^{-9}, N_{\rm iter}=12000$, 
and explore the changes in the energy value for different choices of $N_{\rm samp}$ and $N_{\rm warm}$. 
Since it is computationally costly to explore a wide range of pair values $(N_{\rm samp}, N_{\rm warm})$, 
we picked four choices as representatives: $(40,0)$, $(100,10)$, $(200,20)$, and $(400,40)$.  
The results are visualized in Figure \ref{fig:boxplotN50E619_Nsamp}.
The results show that as $N_{\rm samp}$ and $N_{\rm warm}$ both increase,
the resulting energy increases and gets closer to the results from the  BMZ algorithm.


We also studied the impact of the regularization parameter, $\lambda_{\rm reg}$,
holding fixed $N_{\rm samp}=40, N_{\rm warm}=0, N_{\rm iter}=12000$. 
Results are given in Table \ref{tab:N50E619_sreg}.
Based on the results, we decided to use $\lambda_{\rm reg}=10^{-6}$ for larger size problems.

\begin{table}
\begin{center}
\begin{tabular}{@{}c|cccc@{}}
\toprule
 & 
\multicolumn{3}{c}{Energy value}\\
$\lambda_{\rm reg}$& Mean & Standard deviation & Minimum\\
 \midrule
$10^{-9}$ &-1388.47&	24.81&	-1416.89\\
$10^{-7}$&  -1389.31&	26.85&	-1418.32\\
$10^{-5}$   & -1375.55&	36.13&	-1418.18 \\
\bottomrule
\end{tabular}
\end{center}
\caption{Observed solution energy for the \texttt{RWN50E619} graph with respect to $\lambda_{\rm reg}$, 
the regularization parameter in solving the descent direction. 
\yzcmt{$N_{\rm samp}=40$, $N_{\rm warm}=0$, and $N_{\rm iter}=12000$ for all test cases.}
\yzcmt{See Table \ref{tab:bmz_twographs} for the corresponding results obtained by the BMZ algorithm.}}
\label{tab:N50E619_sreg}
\end{table}

\yzcmt{Table \ref{tab:N50E619_timing} reports the wall-clock time required for the NQS with selected parameter values.  
As we expected, the cost roughly doubles when the value of the parameters including $N_{\rm iter}$,
$N_{\rm samp}$ and $N_{\rm warm}$ doubles.}

\begin{table}[h]
\begin{tabular}{@{}ccc@{}}
\toprule
$N_{\rm iter}$ & $(N_{\rm samp},\, N_{\rm warm})$ & Time in seconds    \\ \midrule
800            & (40,0)                          & $1.5\times 10^{3}$ \\
1600           & (40,0)                          & $2.9\times 10^{3}$ \\
1200           & (40,0)                          & $2.2\times 10^{3}$ \\
1200           & (100,10)                        & $5.2\times 10^{3}$ \\
1200           & (200,20)                        & $1.1\times 10^{4}$ \\
1200           & (400,40)                        & $2.5\times 10^{4}$ \\ \bottomrule
\end{tabular}
\caption{Observed running time in seconds for applying the NQS algorithm to the \texttt{RWN50E619} graph.
$\lambda_{\rm reg}=10^{-9}$ for all cases.}
\label{tab:N50E619_timing}
\end{table}

\subsection{Performance on uniform weight graph }
\label{sec:UWgraph_parameter}

\begin{figure}[h]
\begin{center}
\scalebox{0.58}{\tikzset{every picture/.style={line width=0.75pt}} 


\begin{tikzpicture}[x=0.75pt,y=0.75pt,yscale=-1,xscale=1]
\Large

\draw (388.5,273.75) node  {\includegraphics[width=401.25pt,height=278.63pt]{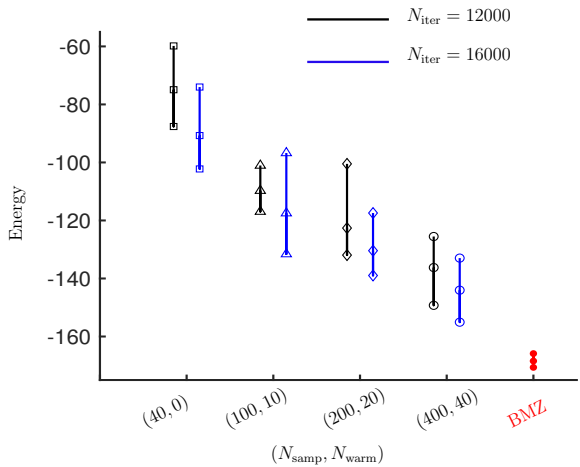}};
\draw [line width=1.5]    (371,105) -- (442,105) ;
\draw [color={rgb, 255:red, 38; green, 12; blue, 236 }  ,draw opacity=1 ][line width=1.5]    (371,142) -- (442,142) ;

\draw (223.15,448.08) node [anchor=north west][inner sep=0.75pt]  [rotate=-333.59]  {$( 40,0)$};
\draw (295.15,449.08) node [anchor=north west][inner sep=0.75pt]  [rotate=-333.59]  {$( 100,10)$};
\draw (380.15,449.08) node [anchor=north west][inner sep=0.75pt]  [rotate=-333.59]  {$( 200,20)$};
\draw (461.15,450.08) node [anchor=north west][inner sep=0.75pt]  [rotate=-333.59]  {$( 400,40)$};
\draw (337.88,474.16) node [anchor=north west][inner sep=0.75pt]  [rotate=-360]  {$( N_{\rm samp} ,N_{\rm warm})$};
\draw (542,450.08) node [anchor=north west][inner sep=0.75pt]  [rotate=-333.59]  {\textcolor{red}{BMZ}};
\draw (456,94) node [anchor=north west][inner sep=0.75pt]    {$N_{\rm iter} =12000$};
\draw (456,128) node [anchor=north west][inner sep=0.75pt]    {$N_{\rm iter} =16000$};
\draw (110.43,299.47) node [anchor=north west][inner sep=0.75pt]  [rotate=-270.14] [align=left] { Energy};

\end{tikzpicture}

}
\caption{Optimal energy for the graph  \texttt{UWN100E246} with respect to $N_{\rm samp}$ and $N_{\rm warm}$ the number of samples and warm samples in Markov chain and $N_{\rm iter}$ the number of reconfiguration steps. Each vertical line corresponds to one choice of the parameters or the BMZ algorithm, and the three markers for each line give the observed minimum, average, and maximum of the energy approximations for 10 runs.  The runs with $N_{\rm iter}=12000$ are in black and those with $N_{\rm iter}=16000$ are in blue. \yzcmt{For the NQS approach, $\lambda_{\rm reg}=10^{-6}$ for all test cases.}}
\label{fig:boxplotN100E246}
\end{center}
\end{figure}

Inspired by the discussion in \cite{Steinerberger2021MaxCutVK}, we also applied the proposed algorithm to uniformly weighted graphs with low edge density.
The idea being that, in this regime, local extrema may not correspond to global extrema, and thus non-local search heuristics may be advantageous compared to local search.
The results for graph \texttt{UWN100E246} are visualized in Figure 
\ref{fig:boxplotN100E246}, using
$\lambda_{\rm reg}=10^{-6}$ and varying the parameters $N_{\rm samp},N_{\rm warm}, N_{\rm iter}$.
The BMZ algorithm produces energy mean of -168.45, standard deviation of 1.52, and minimum of -170.65.
Somewhat surprisingly, the NQS approach fared comparatively worse compared with BMZ in this regime.
Presumably this could be overcome by investing more time and effort into tweaking the parameters and running tests with larger value for $N_{\rm samp}$, $N_{\rm warm}$, and $N_{\rm iter}$.
For example, the current results show consistent improvement in solution optimality as ($N_{\rm samp},N_{\rm warm})$ increases from $(40,0)$
to $(400,40)$. This suggests that simple parameter tweaking,
such as increasing $N_{\rm samp}$ and $N_{\rm warm}$, 
could bring the the NQS results on par with BMZ.



\subsection{Pretrained initialization}
\label{sec:smart_initialization}

\begin{figure}[h]
\begin{center}
\scalebox{0.58}{\tikzset{every picture/.style={line width=0.75pt}} 
\Large
\tikzset{every picture/.style={line width=0.75pt}} 

\tikzset{every picture/.style={line width=0.75pt}} 

\begin{tikzpicture}[x=0.75pt,y=0.75pt,yscale=-1,xscale=1]

\draw (877,263) node  {\includegraphics[width=450pt,height=280pt]{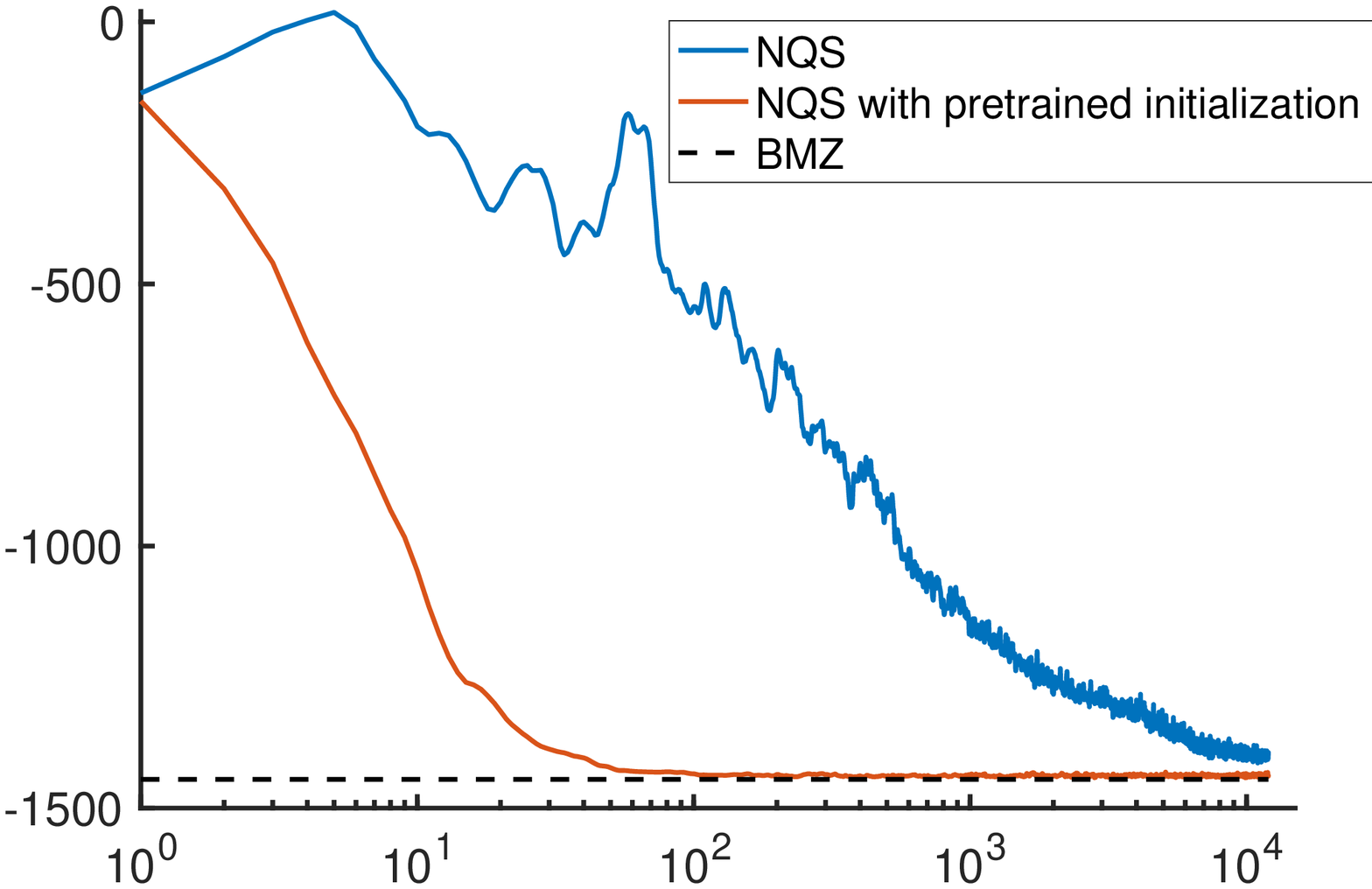}};

\draw (570.43,293.47) node [anchor=north west][inner sep=0.75pt]  [rotate=-270.14] [align=left] { Energy};
\draw (888.88,485.16) node [anchor=north west][inner sep=0.75pt]  [rotate=-359.86]  {$N_{iter}$};

\end{tikzpicture}
%
%
%
%
%

}
\caption{Performance of the proposed NQS algorithm with smart initialization from solution given by the BMZ algorithm applied to the \texttt{RWN50E619} graph with $N_{\rm samp}=40$, $N_{\rm warm}=0$, $N_{\rm iter}=12000$, and \yzcmt{$\lambda_{\rm reg}=10^{-6}$}. }
\label{tab:main_result_initialization}
\end{center}
\end{figure}

The cost of the NQS algorithm can be significantly reduced by initializing the weights and biases of the RBM with an educated guess. If the rotor configuration output from the BMZ algorithm is available,
this can be used to initialize the RBM weights and biases as follows.
Given the output $(\theta_1^\ast,\dots,\theta_n^\ast)$ of BMZ, we initialize the bias on the visible units as
$\vec{c}_i=(r\cos(\theta_i^\ast), r\sin(\theta_i^\ast))$, where $r \geq 0$ is a parameter. The initial hidden bias was chosen to be zero and the weights $a_{ij}$ are randomly drawn from a normal distribution with zero mean and 0.1 standard deviation. 
Figure \ref{tab:main_result_initialization} confirms the intuition that the algorithm converges faster using the pretrained initialization.


\section{Conclusions and future work}
\label{s:conclusions}
 We demonstrated the feasibility of continuous optimization using a quantum-inspired variational algorithm, which we interpret in two ways: either as simulating the continuous-variable QAOA for an efficiently simulable neural-network wavefunction, or alternatively as a realization of continuous-variable natural evolution strategies. Performance was assessed using a challenging non-convex optimization problem and it was found that, despite the non-local nature of the algorithm, a purely local heuristic out-performed the quantum-inspired algorithm in all cases. It is an interesting open problem to find continuous optimization tasks for which variational probabilistic/quantum algorithms offer an advantage compared to local search heuristics. It is also interesting to consider improvements to the choice of classically simulable wavefunction. For example, we have focused on unnormalized densities where samples are generated using the Metropolis-Hastings MCMC algorithm inspired by \cite{carleo2017solving}. A more sophisticated proposal distribution based on Hamiltonian Monte Carlo is expected to deliver significantly improved approximations of Monte Carlo samples. Since the algorithm we described can be also be understood as the endpoint of variational neural annealing (VNA) flow, it will be natural to consider the continuous-variable extension of VNA using normalized densities, and corresponding initialization strategies.

\section{Acknowledgements}
We thank James Stokes (Flatiron Institute) for many helpful discussions. We acknowledge support from the Automotive Research
Center at the University of Michigan (UM) in accordance with Cooperative Agreement W56HZV-19-2-0001 with U.S. Army DEVCOM Ground Vehicle Systems Center. This research was supported in part through computational resources and services provided by UM's Advanced Research Computing.
\nocite{*}
\bibliographystyle{plain}
\bibliography{main}

\appendix
\section{NQS applied to simple graphs}
In this section, we consider a collection of Max-Cut problems defined on simple and/or small graphs
for which the optimal solution can be obtained easily via brute force search. 
The graphs, optimal cut value, and the cut value obtained by the NQS approach coupled with the Procedure-Cut algorithm for converting continuous variable solution to cut are given in Table \ref{tab:simple_results}.
For graphs with 4 nodes, we use $N_{\rm iter}=300$, $N_{\rm samp}=10$, $N_{\rm warm}=0$ and $\lambda_{\rm reg}=10^{-9}$. For graphs with 6 nodes, we use $N_{\rm iter}=1000$, $N_{\rm samp}=40$, $N_{\rm warm}=0$ and $\lambda_{\rm reg}=10^{-9}$.
For graphs with 12 nodes, we use $N_{\rm iter}=4000$, $N_{\rm samp}=40$, $N_{\rm warm}=0$ and $\lambda_{\rm reg}=10^{-9}$.
Note that the observed standard deviation for the 10 runs (each with a different random seed) for all the tests cases is zero. This suggests that for simple graphs with appropriate parameter choices the NQS approach, although being Stochastic in nature, always gives the rotor configurations that leads to the optimal cut value for the graphs.

\begin{table}[h]
\begin{tabular}{@{}c|c|ccc@{}}
\toprule
 Graph                         &                                   & \multicolumn{3}{c}{Cut value via NQS  and Procedure-Cut} \\ \cmidrule(l){3-5} 
 (all edge weight=1)                    & Optimal cut value & Mean        & Standard deviation        & Maximum        \\ \midrule
\includegraphics[width=2.5cm]{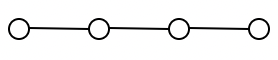}              & 3                            & 3       & 0                     & 3           \\ \midrule
\includegraphics[width=1.2cm]{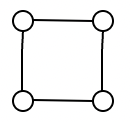}               &    4                               &     4        &                0           &      4          \\ \midrule
\includegraphics[width=1.2cm]{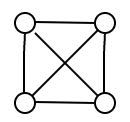}            &      4                             &   4          &       0                    &       4         \\ \midrule
\includegraphics[width=2.5cm]{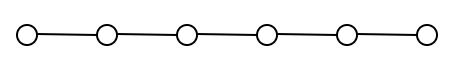}               &     5                              &  5           &                     0      &       5         \\ \midrule
\includegraphics[width=1.2cm]{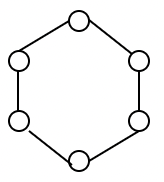}              &       6                            &      6       &                0           &        6        \\ \midrule
\includegraphics[width=0.8cm]{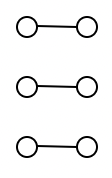} &            3                       &   3          &       0                    &      3          \\ \midrule
\includegraphics[width=1.2cm]{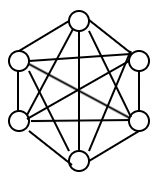}         &        9                           &     9        &                     0      &          9      \\ \midrule
\includegraphics[width=2cm]{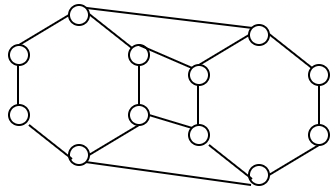}                   &   16                                &      16       &          0                 &         16       \\ \bottomrule
\end{tabular}

\vspace{.5cm}

\caption{Results from applying the NQS and Procedure-Cut algorithm to the Max-Cut problems defined on simple graphs with uniform edge weight. For each problem, the NQS and Procedure-Cut algorithm is run 10 times with 10 different random seeds.  
}
\label{tab:simple_results}
\end{table}
\end{document}